\documentclass[twocolumn,prb,showpacs,citeautoscript,floatfix]{revtex4}
\usepackage{graphicx}

\begin{document}

\title{Two pressure-induced structural phase transitions in TiOCl}

\author{J. Ebad-Allah,$^{1}$ A. Sch\"onleber,$^{2}$ S. van Smaalen,$^{2}$ M. Hanfland,$^{3}$ M. Klemm,$^1$
S. Horn,$^1$ S. Glawion,$^{4}$ M. Sing,$^4$  R. Claessen$^4$, and C. A. Kuntscher$^{1*}$}
\address{
$^1$ Experimentalphysik 2, Universit\"at Augsburg, D-86135 Augsburg, Germany \\
$^2$ Laboratory of Crystallography, Universit\"at Bayreuth, D-95440 Bayreuth, Germany \\
$^3$ European Synchrotron Radiation Facility, BP 220, F-38043 Grenoble, France \\
$^4$ Experimentelle Physik 4, Universit\"at W\"urzburg, D-97074 W\"urzburg, Germany}

\date{\today}

\begin{abstract}
We studied the crystal structure of TiOCl up to pressures of $p$=25~GPa at room temperature     
by x-ray powder diffraction measurements. Two pressure-induced structural phase transitions 
are observed: At $p_{c1}$$\approx$15~GPa emerges an 2$a$$\times$2$b$$\times$$c$
superstructure with $b$-axis unique monoclinic symmetry (space group P2$_1$/$m$). At 
$p_{c2}$$\approx$22~GPa all lattice parameters of the monoclinic phase show a pronounced anomaly.
A fraction of the sample persists in the ambient orthorhombic phase (space group $Pmmn$) over the 
whole pressure range.
\end{abstract}


\maketitle

\section{Introduction}
Recently, high-pressure studies on the low-dimensional Mott-Hubbard insulators TiOCl and TiOBr
have demonstrated the high sensitivity of their electronic and structural properties regarding
the application of pressure.
The first high-pressure experiments on TiOCl and TiOBr single crystals
at room temperature revealed strong changes in the optical response in the infrared frequency range,
with the suppression of the transmittance and increasing optical
conductivity at high pressures.\cite{Kuntscher06,Kuntscher07,Kuntscher10} Based on these results the
possibility of a pressure-induced insulator-to-metal transition was suggested.
However, subsequent electrical transport measurements on powder samples 
did not confirm a metallization
of TiOCl at high pressure and found only an anomaly in the pressure-induced decrease
of the charge gap at $\approx$15~GPa \cite{Forthaus08}.
A controversy regarding the high-pressure electronic properties of TiOCl also exists from the theoretical
point of view: {\it Ab initio} calculations were carried out based on density functional theory
within the LDA+U approximation \cite{Blanco-Canosa09} and using Car-Parrinello molecular dynamics
\cite{Zhang08}. While the former claims an insulating phase up to at least 30~GPa, the latter predicts
the metallic character of the high-pressure phases.

According to both experiment and theory, TiOCl exhibits structural phase transitions under pressure 
at room temperature. Signatures of a structural phase transition in TiOCl and TiOBr 
under pressure were first observed in x-ray powder diffraction studies.\cite{Kuntscher07,Kuntscher08} 
This structural phase transition coincides with an anomaly in the pressure-dependent electrical
and optical properties at $p_{c1}$$\approx$15~GPa.\cite{Kuntscher07,Kuntscher08,Forthaus08}
Blanco-Canosa et al.~\cite{Blanco-Canosa09} confirmed the occurrence of a structural phase transition,
but the corresponding critical pressure was considerably lower ($\approx$10~GPa). The high-pressure
phase was specified as $b$-axis unique monoclinic $P2_1/m$ phase with a strong dimerization of the 
1D spin chain along $b$.\cite{Blanco-Canosa09} It was furthermore claimed \cite{Blanco-Canosa09} 
that the dimerization is of pure electronic origin, i.e., resembles that of a conventional Peierls 
insulator, and that the magnetic interaction plays only 
a minor role at high pressure.
It is noticed here that the reported
symmetry\cite{Blanco-Canosa09} is 
not a subgroup of the ambient $Pmmn$ symmetry. 
The high-pressure phase thus is not a simple superstructure 
of the structure at ambient conditions, 
but it requires major rearrangements 
affecting the connectivity between atoms.
These might be provided by the observed 
monoclinic angle of $\beta$${\sim}\,99$$^{\circ}$ as
compared to 90$^{\circ}$ in the orthorhombic phase. 
The reported atomic positions at 
$2e$:\ $(\frac{1}{4}, y, z)$ would require a 
mirror plane perpendicular to the 
$a$-axis and thus are incompatible 
with the reported monoclinic 
$b$-axis.\cite{Blanco-Canosa09} 
Besides, several aspects of the high-pressure phases still need to be clarified, e.g., the values 
of the critical pressures and possible superstructures being induced at higher pressures.

Here we present x-ray diffraction data on TiOCl for an extended pressure range up to $\sim$25~GPa. 
Our goal is to clarify the inconsistencies regarding the experimental findings for the high-pressure
crystal structure of TiOCl.
Besides a transition from the orthorhombic $Pmmn$ to the monoclinic $P2_1/m$ crystal structure with
an $2a$$\times$2$b$$\times$$c$ superstructure, we find a pressure-induced isostructural phase 
transition for the monoclinic phase with anomalies in the lattice parameters.

\section{Experiment}
\label{sectionexperiment} 

Single crystals of TiOCl were synthesized by gas transport from TiCl$_3$ and TiO$_2$.\cite{Schaefer58}
TiOCl crystallizes in the space group $Pmmn$ at ambient conditions and consists of
distorted TiO$_4$Cl$_2$ octahedra.
Pressure-dependent x-ray powder diffraction measurements at room temperature
were carried out with monochromatic radiation ($\lambda$= 0.4128~\AA) at beamline
ID09A of the European Synchrotron Radiation Facility (ESRF) at Grenoble.
Crystals were gently ground and placed into a diamond anvil cell (DAC). The material of the gasket
was stainless steel, and its initial thickness and hole diameter was 40~$\mu$$m$ and 
150~$\mu$$m$, respectively.
The applied pressures $p$ were determined with the ruby fluorescence method.\cite{Mao86}
Helium served as hydrostatic pressure-transmitting medium. Diffraction patterns were recorded 
with an image plate detector and then integrated with FIT2D\cite{Hammersley98} to yield 
intensity vs 2$\theta$ diagrams. The DAC was rotated by $\pm$3$^\circ$ during the exposure to
improve the powder averaging. We carried out LeBail fits of the diffraction data using
the Jana2006 software,\cite{Petricek06} in order to determine the lattice parameters as a function
of pressure. Rietveld refinements of the diffraction data could not be carried out due to the 
preferred orientation of the crystallites inside the DAC, as described earlier.\cite{Kuntscher08}

\begin{figure}[t]
\includegraphics[width=1\columnwidth]{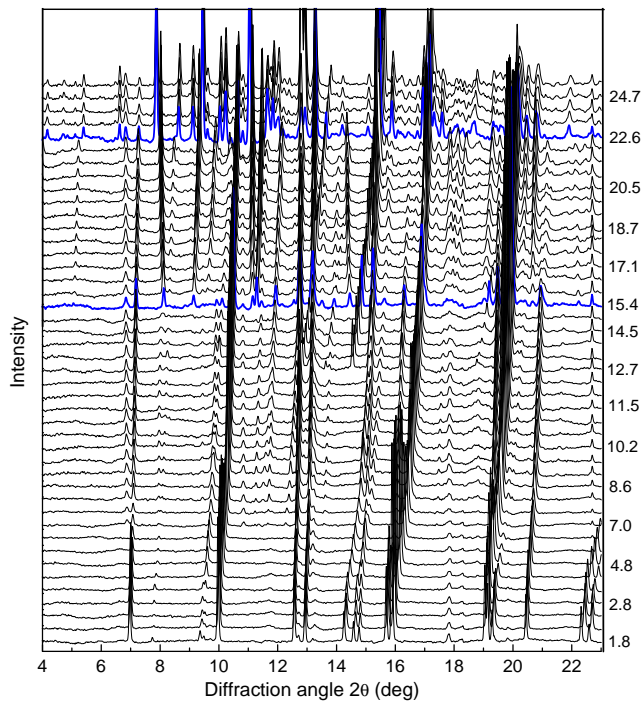}
\caption{Room-temperature x-ray powder diffraction diagrams of TiOCl at high pressures.
The numbers on the right, vertical axis indicate the applied pressures in GPa.
The diffraction diagrams at the critical pressures p$_{c1}$=15.4~GPa and p$_{c2}$=22.6~GPa
are highlighted by bold, blue lines.}
\label{fig:overview}
\end{figure}

\begin{figure}[t]
\includegraphics[width=0.95\columnwidth]{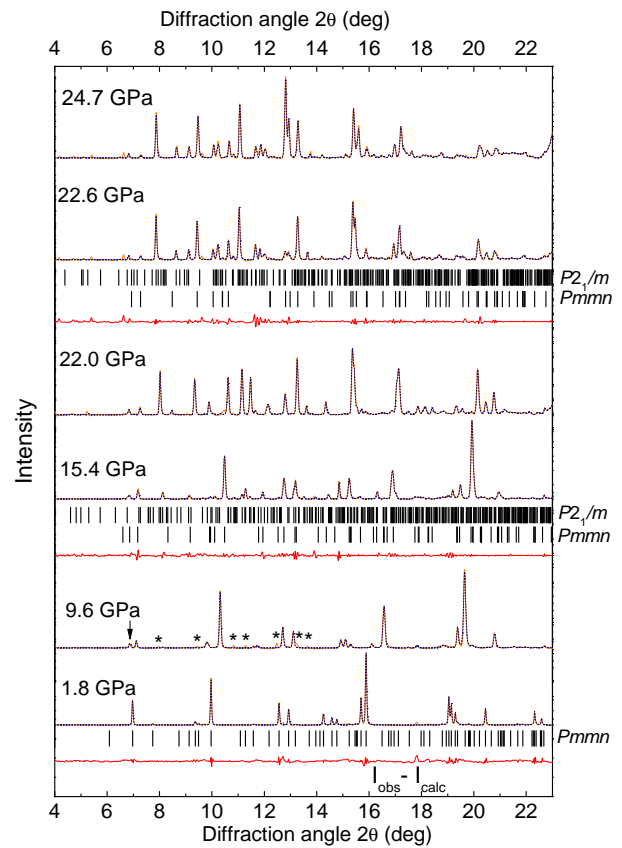}
\caption{Room-temperature x-ray powder diffraction diagrams (full, orange lines) of TiOCl at high pressures 
together with the LeBail fits (dotted, blue lines).
For the pressures 1.8, 15.4, and 22.6~GPa) the difference curve ($I_{obs}-I_{calc}$) between
the diffraction diagram and the LeBail fit is shown. Markers show the calculated peak positions
for the various phases. The arrow indicates the reflection from the ruby chip.
The asterisks mark the reflections due to the monoclinic phase already present above $\sim$7~GPa.}
\label{fig:diffraction}
\end{figure}

\begin{figure}[t]
\includegraphics[width=0.9\columnwidth]{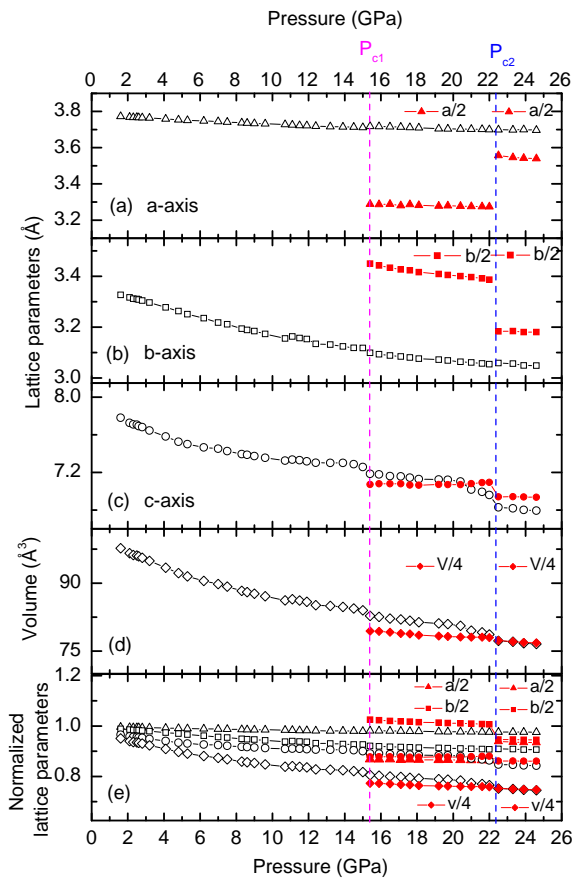}
\caption{(Color online) Results of the LeBail fits of the room-temperature x-ray diffraction diagrams 
of TiOCl: lattice parameters and unit cell volume as a function of pressure 
[$Pmmn$: black, open symbols; $P$2$_1$/$m$: red (gray), 
full symbols]. Lines are guides to the eye. For the normalization of the lattice parameters and the unit
cell volume the values of Ref.~\onlinecite{Sasaki05} were used.}
\label{fig:parameters}
\end{figure}

\begin{figure}[t]
\includegraphics[width=0.75\columnwidth]{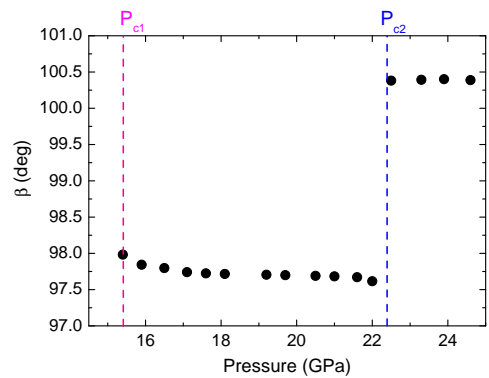}
\caption{Angle $\beta$ of the unit cell of the $P$2$_1$/$m$ monoclinic crystal structure at room temperature
as a function of pressure.}
\label{fig:beta}
\end{figure}

\section{Results and discussion}
\label{sectionresults}

Fig.~\ref{fig:overview} shows the diffraction diagrams of TiOCl for all measured pressures. 
Fundamental changes in the diffraction diagrams are observed at 15.4 and 22.6 GPa, which define
the critical pressures p$_{c1}$ and p$_{c2}$ for the two structural phase transitions.
The x-ray powder diffraction diagrams of TiOCl for selected pressures up to $\sim$25~GPa are presented
in Fig.~\ref{fig:diffraction}. Up to the critical pressure $p_{c1}$$\approx$15~GPa the diffraction diagrams
can be described by LeBail fits applying symmetry and lattice parameters of the
ambient-pressure, orthorhombic crystal structure (space group $Pmmn$) and assuming contributions
from the ruby chips added for the pressure determination. 
The lattice parameters and unit cell volume of the orthorhombic crystal structure obtained 
by the LeBail fits are shown in Fig.\ \ref{fig:parameters} and are in agreement with those 
reported in Refs.~[\onlinecite{Kuntscher08,Kuntscher09}].
We find a slightly nonlinear decrease of the lattice parameters with increasing pressure.

Above $P_{c1}$$\approx$15~GPa the diffraction diagrams can no longer be described by a single phase,
but a good fit of the data can be achieved by assuming the coexistence of two phases --
namely an orthorhombic phase (space group $Pmmn$) and a monoclinic phase (space group $P$2$_1$/$m$, 
$b$-axis unique) with double $a$- and $b$-axes.
The lattice parameters and the angle $\beta$ of the monoclinic unit cell hardly change up to $\sim$22~GPa 
(see Figs.~\ref{fig:parameters} and \ref{fig:beta}).
The coexistence of two phases extends over a broad pressure range and signals the
sluggish character of the phase transition. A large pressure range of phase coexistence,
of 10~GPa or larger, has been observed for several first-order structural phase transitions
 \cite{Rozenberg09,Loa04,Xu01} and
suggests that the two phases are almost energetically degenerate.\cite{comment1}

It should be noted that already above $\sim$7~GPa several weak reflections occur (marked with
asterisks in Fig.~\ref{fig:diffraction}, which
cannot be related to the orthorhombic phase\cite{comment2} but to the monoclinic phase, which 
fully develops at p$_{c1}$ as described above. The appearance of diffraction peaks related 
to the monoclinic phase at pressures lower than p$_{c1}$ might be due to non-hydrostatic 
conditions in the DAC, leading to locally higher pressures than indicated by the ruby 
fluorescence.

The appearance of additional reflections above p$_{c1}$
due to a mono\-clinic phase {\it in addition} to those of the
orthorhombic phase is also observed in the diffraction diagrams of Ref.~\onlinecite{Blanco-Canosa09}.
There, the additional reflections were attributed to a monoclinic phase with a doubling of the unit 
cell along the $b$-axis. Furthermore, it was claimed\cite{Blanco-Canosa09} that the mono\-clinic phase 
above 10~GPa at room temperature resembles the monoclinic spin-Peierls phase occurring below the critical 
temperature $T_{SP}$ at ambient pressure. This led to the conclusion that $T_{SP}$ increases with 
increasing pressure, and it was proposed that $T_{SP}$ reaches room temperature for pressures above 
10~GPa.\cite{Blanco-Canosa09}
In contrast to the results of Blanco-Canosa {\it et al.}\cite{Blanco-Canosa09} we find a doubling of 
the unit cell along both $a$- and $b$-axes in the monoclinic cell. This discrepancy could be explained
by the fact that the results of Ref.~\onlinecite{Blanco-Canosa09} were restricted to the pressure 
range $\leq$15.2~GPa, where the monoclinic phase is not yet fully developed (which is the case only 
above p$_{c1}$=15.4~GPa according to our data). Furthermore, it is important to stress the differences
between the crystal structures of the ambient-pressure spin-Peierls phase at low
temperatures and the high-pressure dimerized phase at room-temperature: While the former shows
an $a$-axis unique monoclinic symmetry with a monoclinic angle 
$\alpha$${\sim}\,90$$^{\circ}$,\cite{Shaz05,Schoenleber06,Schoenleber08} the latter has a $b$-axis unique
mono\-clinic symmetry with monoclinic angle $\beta$${\sim}\,99$$^{\circ}$.\cite{Blanco-Canosa09} 

It is interesting to note that an approximated doubling of the monoclinic unit cell along the $a$-axis 
was recently observed in a high-pressure x-ray diffraction study on TiOCl at $T$=6~K:\cite{Prodi09} 
Starting from the ambient-pressure, low-temperature monoclinic spin-Peierls phase, 
Prodi et al.\ found a pressure-induced suppression of the dimerization along the $b$-axis in the vici\-nity 
of a first-order structural phase transition at around 13~GPa \cite{Prodi09}. The high-pressure phase 
shows an incommensurate superstructure of the type (2$a$-$\epsilon$)$\times$$b$$\times$$c$ and might 
be interpreted in terms of a conventional Peierls state.

\begin{figure}[t]
\includegraphics[width=0.85\columnwidth]{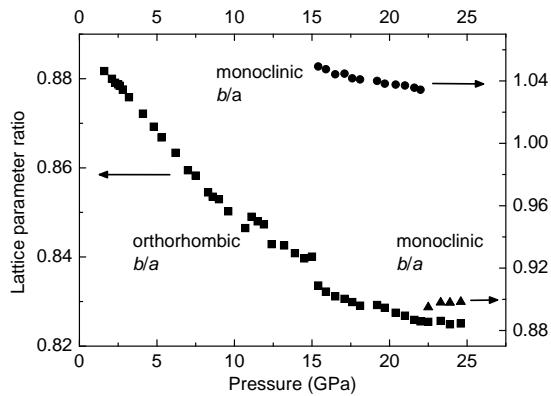}
\caption{Lattice parameter ratio as a function of pressure: Ratio $b$/$a$ for the orthorhombic 
phase ($Pmmn$, squares) for the whole studied pressure range; ratio $b$/$a$ for the 
monoclinic phase ($P$2$_1$/$m$, circles) in the pressure range $p_{c1}$$<$$p$$<$$p_{c2}$; 
ratio $b$/$a$ for the monoclinic phase ($P$2$_1$/$m$, triangles) 
in the pressure range $p$$>$$p_{c2}$.}
\label{fig:ratio}
\end{figure}

At $p_{c2}$$\approx$22~GPa again significant changes in the x-ray diffraction diagram
are observed (see Figs.~\ref{fig:overview} and \ref{fig:diffraction}), indicating the occurrence 
of a second structural phase transition: All lattice parameters of the monoclinic phase show 
pronounced anomalies at $p_{c2}$, as seen in Figs.~\ref{fig:parameters} and \ref{fig:beta}. 
Above $p_{c2}$ several weak reflections appear at diffraction angles below 10$^{\circ}$, 
which can neither be related to the orthorhombic and monoclinic phase, nor explained in 
terms of reflections from the ruby chips, the diamond anvils or the gasket. In view of the 
complicated phase diagram of TiOCl under pressure, with a coexistence of multiple phases, 
we explain these weak reflections in terms of a third phase appearing above $p_{c2}$.
In order to specify the structural details of this third phase, x-ray diffraction data
at higher pressures are needed.

In the following we speculate on the possible mechanism driving the structural phase transitions
in TiOCl under pressure. We propose that their occurrence  
is closely linked to the tuning of the anisotropy or dimensionality of the system under pressure.
For an illustration of the dimensionality, we plot and compare in Fig.~\ref{fig:ratio} the lattice 
parameter ratios for the various phases. Here we assume that the average Ti-Ti distance scales with the lattice 
parameters. For the orthorhombic phase ($Pmmn$) the ratio $b$/$a$ is shown for the whole pressure range studied. 
For the high-pressure phases we take into account the superstructures and thus plot the ratio $b$/$a$
for the monoclinic phase ($P$2$_1$/$m$) in the pressure range $p$$>$$p_{c1}$.

For the orthorhombic phase the ratio $b$/$a$ monoto\-nically decreases with increasing pressure
for the whole pressure range studied, with the tendency to saturation at high pressures.
This indicates a pressure-induced enhancement of the one-dimensional character of the system consistent with
an earlier report \cite{Blanco-Canosa09}. The enhanced one-dimensional character under pressure is due to the 
larger compressibility of the lattice along the $b$-direction.\cite{Blanco-Canosa09} 
At $p_{c1}$ a monoclinic phase with doubled unit cell along $a$ and $b$ is energetically degenerate
with the orthorhombic phase. For this monoclinic phase we find a ratio $b$/$a$ close
to one, i.e., the system is {\it close to two-dimensional}, up to $p_{c2}$.
This is due to the significant shrinkage of the unit cell along the $a$-direction and its
enlargement along the $b$-direction,
compared to the orthorhombic phase (Fig.~\ref{fig:parameters}). Accordingly,
the interchain interaction is expected to play a major role in determining the electronic and 
magnetic properties of TiOCl in the mono\-clinic high-pressure phase. 
The importance of the interchain interaction in TiOCl at ambient pressure is
commonly accepted.\cite{Ruckamp05,Fausti07,Zhan08} Our data suggest an enhanced interchain 
interaction, comparable to the intrachain interaction, for $p_{c1}$$<$$p$$<$$p_{c2}$. 
Above $p_{c2}$ the ratio $b$/$a$ drops to $\approx$0.9, i.e., similar to the ratio $b$/$a$
of the orthorhombic phase at ambient conditions. Thus the material becomes more one-dimensional
above $p_{c2}$.

The occurrence of two pressure-induced phase transitions in TiOCl was predicted 
by Zhang et al.~[\onlinecite{Zhang08}]: At $p_{c1}$ a structural phase transition with a change 
from an orthorhombic ($Pmmn$) to a monoclinic crystal structure ($P$2$_1$/$m$) accompanied by a 
doubling of the unit cell along $b$ is expected.
The concomitant insulator-to-metal transition is proposed to be due to a broadening of the 
electronic bands and a redistribution of electronic occupation among the three $t_{2g}$ 
bands.\cite{Zhang08} A second first-order structural phase
transition  from the dimerized monoclinic $P$2$_1$/$m$ phase to a uniform (undimerized) metallic phase
with orthorhombic $Pmmn$ symmetry is predicted at the critical pressure
$p$=$p_{c2}$$\approx$1.26$\cdot$$p_{c1}$.\cite{Zhang08}
Because of the structural changes and the orbital repopulation a dimensional crossover of TiOCl 
from quasi-one-dimensional to quasi-two-dimensional was proposed to occur above $p_{c2}$.\cite{Zhang08}
A tendency of the system at high pressures towards two-dimensionality is consistent with our 
experimental results. However, it occurs already above $p_{c1}$ and is limited to the pressure
range $p_{c1}$$<$$p$$<$$p_{c2}$ according to our data. Discrepancies also exist regarding the 
doubling of the unit cell along the $a$ axis, and, furthermore, we do not find a pure undimerized, 
orthorhombic phase at high pressures, as theory\cite{Zhang08} predicts.

Interestingly, the pressure-induced structural instabilities seem to 
strongly depend on temperature: While at low temperature the dimerization along the $b$-axis
becomes energetically unfavorable above $\sim$13~GPa and a switching from the 
spin-Peierls phase to a phase with an incommensurate charge density wave along the $a$-axis
occurs,\cite{Prodi09} at room temperature the dimerization along $b$ for $p$$>$$p_{c1}$
persists at least up to 25~GPa according to our findings. Possible incommensurabilities 
of the superstructures at room temperature remain to be clarified.

\section{Conclusions}
\label{summary}

Pressure-dependent x-ray powder diffraction data show the occurrence of two
structural phase transitions in TiOCl at room temperature. Above $p_{c1}$$\approx$15~GPa a 
monoclinic phase (space group $P$2$_1$/$m$, $b$-axis unique) with a dimerization along the 
$a$- and $b$-direction emerges. At $p_{c2}$$\approx$22~GPa all lattice parameters of the 
monoclinic phase show pronounced anomalies. A fraction of the sample persists in the ambient 
orthorhombic phase (space group $Pmmn$) over the whole pressure range studied.

\subsection*{Acknowledgements}

We acknowledge the ESRF Grenoble for the provision of beamtime.
Financial support by the DFG, including the Emmy Noether-program, SFB 484, DFG-CL124/6-1,
and Sm55/15, is acknowledged.


\begin{thebibliography}{10}
\bibitem{comment1}
Please note that the volume fractions of the phases cannot be extracted based on the
LeBail fits.
\bibitem{comment2}
The appearance of reflections above a certain pressure can be related to the loss of 
the preferred orientation of the crystallites in the DAC, as for example observed 
in the 12.7~GPa diffraction diagram, where a new reflection related to the orthorhombic phase 
is observed between 2$\theta$=14$^{\circ}$ and 15$^{\circ}$.
\end{thebibliography}
\end{document}